\documentclass[epj]{svjour}
\usepackage{graphics}
\usepackage{epstopdf}

\newcommand{\Npart}{$\langle N_{part}\rangle$}

\newcommand{\sNN}[1]{$\sqrt{s_{NN}} = #1$ GeV}
\newcommand{\Lam}{$\Lambda $ }
\newcommand{\ALam}{$\bar{\Lambda} $ }
\newcommand{\Kaon}{$K^{0}_{S} $ }
\newcommand{\X}{$\Xi^{-}$}
\newcommand{\AX}{$\bar{\Xi}^{+}$}
\newcommand{\Omeg}{$\Omega^{-}$}
\newcommand{\AOmeg}{$\bar{\Omega}^{+}$}
\begin{document}
\title{Strange Particle Production at RHIC}
\author{Anthony R. Timmins\inst{1,2} for the STAR Collaboration}                     
\institute{Wayne State University, Department of Physics and Astronomy, 666 W. Hancock, Detroit, MI 48201, USA \and \email{tone421@rcf.rhic.bnl.gov}}

\offprints{}          
%
%
\date{Received: date / Revised version: date}
\abstract{
We report STAR measurements of mid-rapidity yields for the \Lam, \ALam, \Kaon, \X, \AX, \Omeg, \AOmeg particles in Cu+Cu and Au+Au \sNN{200} collisions. We show that at a given number of participating nucleons, bulk strangeness production is higher in Cu+Cu collisions compared to Au+Au collisions at the same center of mass energy, counter to predictions from the Canonical formalism. We compare both the Cu+Cu and Au+Au yields to AMPT and EPOS predictions, and find they reproduce key qualitative aspects of the data. Finally, we investigate other scaling parameters and find bulk strangeness production for both the measured data and theoretical predictions, scales better with the number participants that undergo more than one collision.
\PACS{
      {PACS-key}{discribing text of that key}   \and
      {PACS-key}{discribing text of that key}
     } 
} 
\maketitle
\section{Introduction}
\label{intro}
Measurements of strangeness production in heavy-ion collisions were originally conceived to be the smoking gun of QGP formation \cite{RafalMull}. It was argued that due to a drop in the strange quark's dynamical mass and increased production cross section, strangeness in the QGP would equilibrate on small time scales relative to a hadronic gas. Assuming a thermally equilibrated QGP hadronizes into a maximum entropy state, this hypothesis can be tested for heavy-ion collisions by comparing final state strangeness yields per participant to thermal model predictions from the Canonical formalism \cite{CanoncialSuppression}. These predictions have qualitatively reproduced various aspects of the data from Au+Au \sNN{200} collisions at RHIC, however, as with SPS energies, a complete theoretical description has yet to be achieved \cite{HelensPRC}. Furthermore, the strangeness saturation factor, $\gamma_{S}$, which characterizes the deviation in strangeness yields from thermal expectations, has been shown to approach unity in central Au+Au \sNN{200} collisions \cite{AuAuLambda}. In these proceedings, we review the previously measured Au+Au \sNN{200} data \cite{AuAuLambda}\cite{AuAuK0S} in the context of new data from approximately 40 million Cu+Cu \sNN{200} collisions recorded at the Relativistic Heavy Ion Collider in 2005. Measurements at AGS showed $K^{+}$ and $K^{-}$ yields to be higher in lighter systems compared to the respective values in heavy systems at a given number of participants \cite{AGS}, while measurements at the SPS showed higher $K/ \pi$ ratios for the light systems also at a given number participants \cite{SPS}. Whether these trends continue up to RHIC energies, and what new information can be learned from strangeness production as QGP signature at RHIC, will be central issues for this analysis.
\section{Yield Extraction}
\label{ana}
The STAR Time Projection Chamber (TPC) is used to extract yields of  \Lam, \ALam, \Kaon, \X, \AX, \Omeg, and \AOmeg  as a function of transverse momentum, $p_T$,  via their dominant weak decay channels, which are $\Lambda \rightarrow p^{+}+\pi^{-}$, $\bar{\Lambda} \rightarrow p^{-}+\pi^{+}$, $K^{0}_{S} \rightarrow \pi^{+}+\pi^{-}$, $\Xi \rightarrow \Lambda+\pi^{-}$, $\bar{\Xi} \rightarrow \bar{\Lambda}+\pi^{+}$, $\Omega \rightarrow \Lambda+K^{-}$, and $\bar{\Omega} \rightarrow \bar{\Lambda}+K^{+}$ respectively. The decay products enter the TPC and are reconstructed using STAR's tracking software. The raw particle yields are calculated from the respective invariant mass calculations for the V0 and cascade candidates, and a combination of topological, energy loss, and kinematic restrictions are placed to ensure the combinatorial background is minimal and can be described with a $2^{nd}$ order polynomial. To calculate the reconstruction efficiency, Monte Carlo particles are generated and propagated through a GEANT detector simulation. The generated charge clusters are then embedded into raw data, and the normal reconstruction process is applied. The efficiency is then defined as the ratio of reconstructed particles to the number of generated particles, subject to the previously mentioned restrictions on the raw data. The efficiencies are applied to the raw data to give the corrected spectra. The $ $\Lam and $ $\ALam yields have contributions from the weak decay of charged and neutral $\Xi$ (anti) particles subtracted. Systematic uncertainty on the spectra points was found to be due to 1) An improper treatment of the vertex resolution in simulation, 2) Slight mismatch of the raw and simulated TPC hit distributions, and 3) Run-day variations in the yield due to variations in the TPC gain. Finally, we chose either an $m_T$ exponential or Maxwell-Boltzman in order to extrapolate to the low $p_T$ region beyond the TPC acceptance so that $dN/dy$ can be calculated for a given particle. More detailed descriptions of the yield extraction can be found here \cite{Mythesis}\cite{STARpp}. 
\begin{figure*}[t]
\resizebox{1\textwidth}{!}{%
  \includegraphics{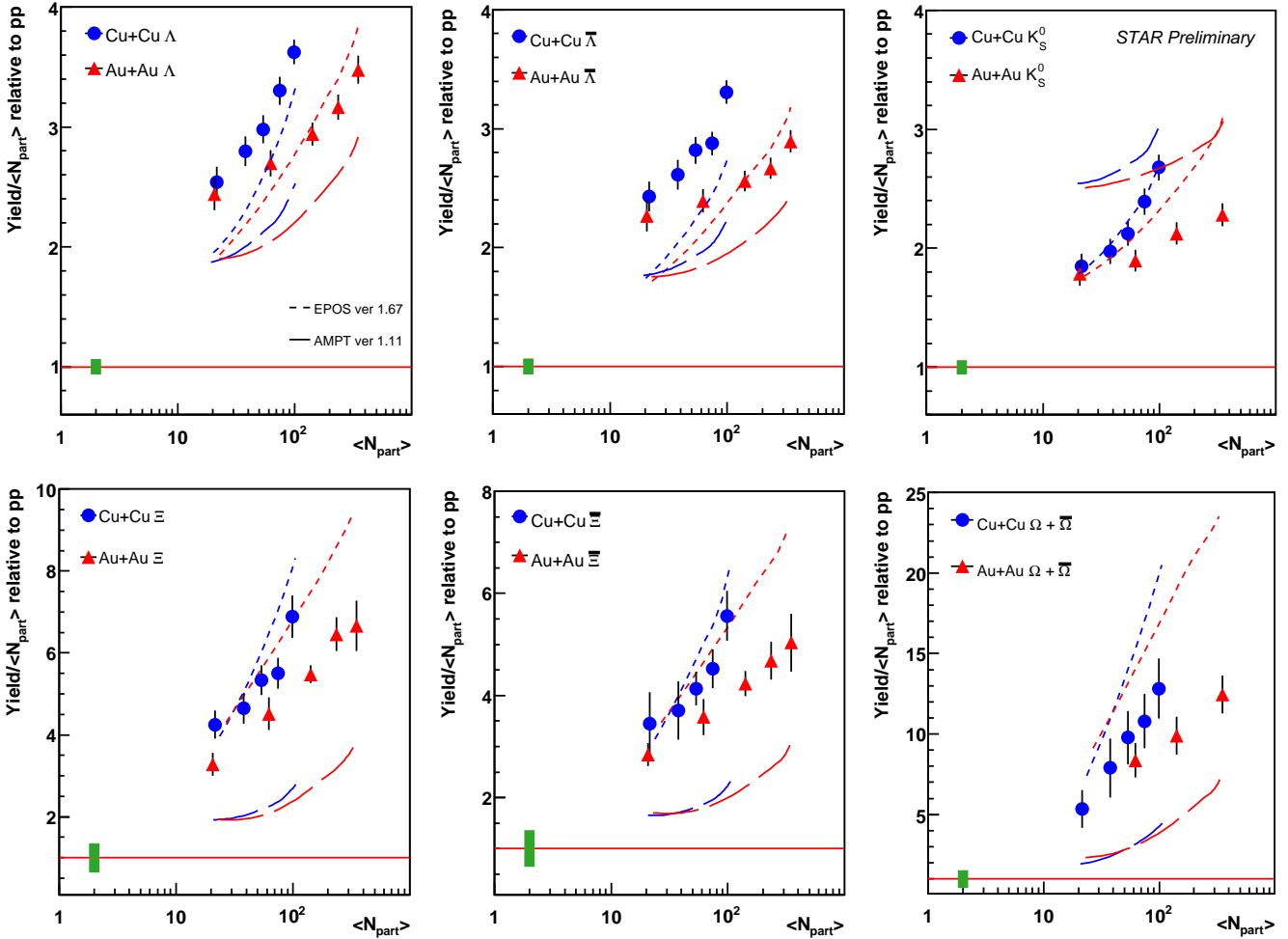}
 }
\caption{Preliminary mid-rapidity per participant yields ($dN/dy$ per \Npart) of singly and multi strange particles for Cu+Cu and Au+Au collisions with \sNN{200} divided by the respective p+p values. The $\Lambda$ and $\bar{\Lambda}$ yields have been feed down subtracted in all cases. The green bars show the normalization uncertainties, and the uncertainties for the heavy-ion points are the combined statistical and systematic. The dotted and dashed lines show EPOS and AMPT predictions respectively where the p+p reference corresponds to the experimentally measured yields. Modulo the $\Omega$ particles, the Cu+Cu trends (blue) in all cases lie above Au+Au (red) for a given particle species and model.}
\label{fig:1}       
\end{figure*}

\section{Participant Scaled Yields}
\label{sec:1}

Figure \ref{fig:1} shows participant scaled strangeness yields relative to p+p values for Cu+Cu and Au+Au collisions with \sNN{200}. This variable is also known as the \emph{enhancement factor} as it characterizes the deviation from p+p collisions for participant scaled heavy-ion yields. The top panels show the singly strange particles which carry the majority of the strangeness produced, while the bottom panels show multi strange particles. For the singly strange particles it is clear that, for a given system, rises in participant yields are observed for each particle which is predicted by the Canonical formalism. However, at a given number participants above approximately 60 for Cu+Cu collisions, higher yields are observed. This is inconsistent with the Canonical formalism as it predicts a unique value for the enhancement factor at a given number of participants assuming the volume of the system is proportional to number of participants. Such a prediction also relies on a constant baryon chemical potential and constant chemical freeze out temperature for both systems which has been shown to be the case for Cu+Cu and Au+Au \sNN{200} collisions within experimental uncertainty  \cite{AnetaPri}. Similar patterns are observed in the multi strange sector although any actual differences in Cu+Cu and Au+Au yields at a given number of participants are not as clear due to the larger fractional uncertainties for those measurements. 

\begin{figure}[h]
\resizebox{0.5\textwidth}{!}{%
  \includegraphics{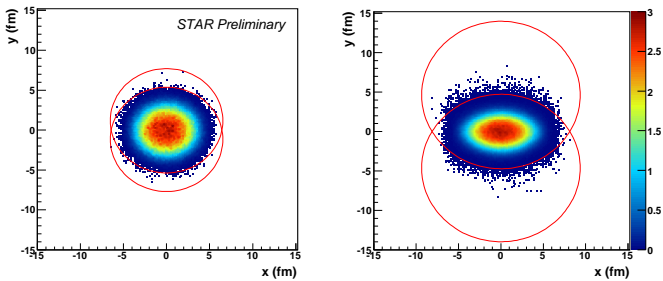}
}
\caption{Preliminary Monte Carlo Glauber calculation of participant densities (fm$^{-2}$) in the transverse plane for central Cu+Cu collisions (left) and mid-central Au+Au collisions (right) with \sNN{200}. The impact parameters for Cu+Cu and Au+Au are fixed to 2.37 and 9.27 fm respectively, and were chosen so both systems yield \Npart$\simeq$99. The radii at the redlines contain $99.7\%$ of the respective Wood-Saxon distributions.}
\label{fig:2}       
\end{figure}

We also make comparisons to predictions from the AMPT \cite{AMPT} and EPOS \cite{EPOS} models which have been shown to describe the bulk features of hadron production well for Au+Au \sNN{200} collisions. The AMPT model is based on HIJING, and thus describes particle production in heavy-ion collisions via intra-nucleon string excitation and breaking (soft), and mini-jet fragmentation (hard) where the excited nucleons fragment independently. For the settings used, mini-jet partons and newly produced hadrons can re-scatter, and without these mechanisms, AMPT results reduce to that of HIJING. The EPOS model describes particle production with core and corona contributions. The core occupies the high participant density of collision zone (see figure  \ref{fig:2}) and aims to mimic various QGP behavior. Once formed, it expands then hadronizes at a critical energy density similar to values predicted by Lattice QCD. The hadronization is treated via a statistical framework where strangeness is over-saturated with $\gamma_{S}=1.3$. This was chosen to the fit the Au+Au \sNN{200} data by the EPOS authors. Corona production occurs in the low density region and can be thought of as a superposition of p+p collisions where strangeness is under-saturated at RHIC energies \cite{STARpp}. From peripheral (60-80\%) Au+Au \sNN{200} collisions up to the most central, it was shown previously that core production is the major source of strangeness production where the relative contribution to total strangeness production rises from $\sim65\%$ to $\sim100\%$ \cite{EPOS}. Finally, for a given model, the default parameter set is used for all systems, centralities and particle species.
For the singly strange particles, both the AMPT and EPOS models reproduce three key qualitative aspects the data: rises in per participant yields for a given system, higher yields at a given number of participants for mid-central and central Cu+Cu collisions compared to Au+Au, and a merging in per participant yields for peripheral Cu+Cu  and very peripheral Au+Au collisions. Quantitatively, EPOS is nearly always closer to the data than AMPT. With regard to the multi strange particles, both AMPT and EPOS also predict the previously mentioned trends which are consistent with the data. EPOS is again closer to the Cu+Cu yields for all particles and over predicts the Au+Au yields, while AMPT under predicts both Cu+Cu and Au+Au yields. 

In figure \ref{fig:2} we show probability distributions for participant densities in the transverse plane for Cu+Cu and Au+Au \sNN{200} collisions where the mean number of participants are roughly the same in each case. The densities are clearly distributed differently owing to the different spatial geometries for the Cu and Au nucleons. In order to gain further insight into qualitative trends for strangeness production, the next section investigates the effects of geometry on the relations between other Glauber calculated quantities and strangeness yields for both the measured data and theoretical predictions. 

\section{Empirical Scalings}
\label{sec:2}
 
Given the similar systematic behaviour of all particles in figure \ref{fig:1}, we calculate the following quantity in order to concisely analyze strangeness production under various scaling hypotheses:
\begin{equation}
\label{equ:total}
\frac{d N_{\langle s+\bar{s} \rangle}}{dy} = 1.48\frac{d N_{\Lambda}}{dy}+4 \frac{dN_{K^{0}_{S}}}{dy}+1.48\frac{d N_{\bar{\Lambda}}}{dy} + 4\frac{d N_{\Xi}}{dy}+4\frac{d N_{\bar{\Xi}}}{dy} 
\end{equation}
which aims to approximate total strangeness production at mid-rapdity. The factors of 1.48 account for the production of (anti) $\Sigma$ particles which have a similar quark content to the $\Lambda$ particles, the factors of 4 accounts for the production of $K^{\pm}$, $K^{0}_L$, charged and neutral $\Xi$ particles \cite{RafaleskiTextboo}. The $\Omega$ particles are neglected due to incomplete centrality coverage in Au+Au \sNN{200} collisions, and they make a negligible contribution to the total $s+\bar{s}$ yield (less than $1\%$ for central Au+Au). 
 \begin{figure*}
\resizebox{1\textwidth}{!}{%
  \includegraphics{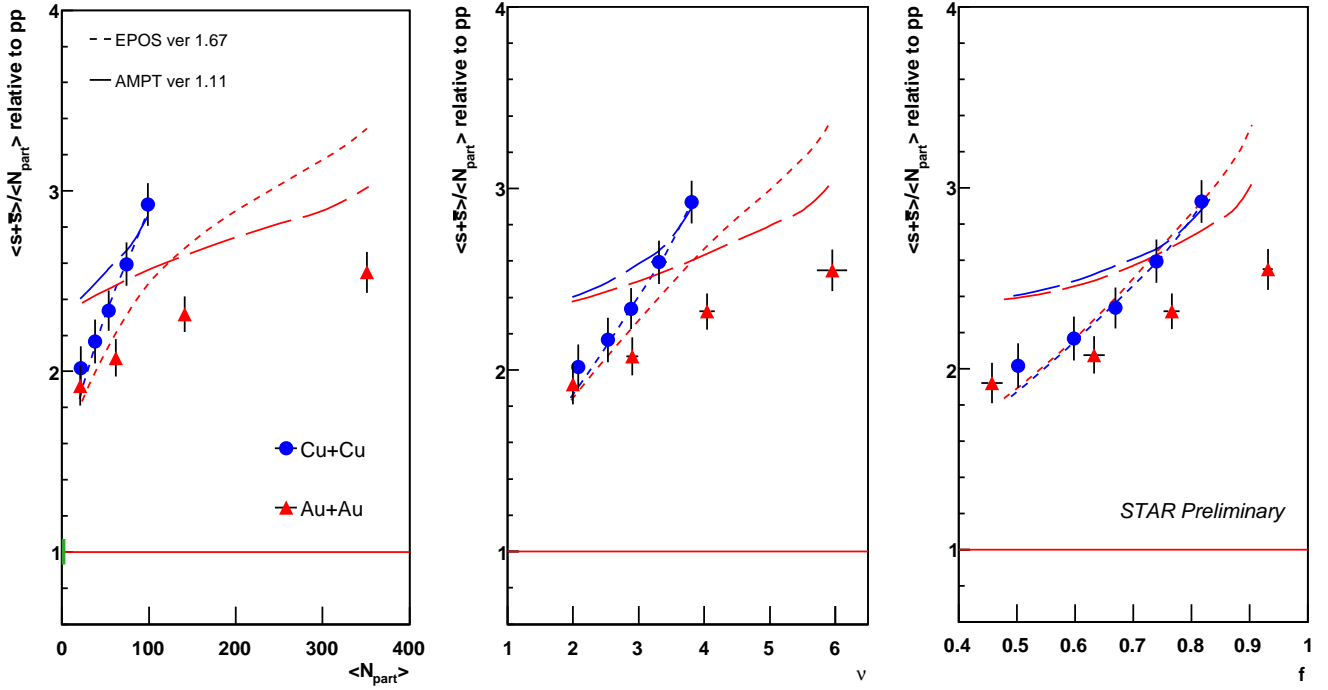}
}
\caption{Preliminary mid-rapidity $dN/dy$ for $s+\bar{s}$ per participant relative to p+p as a function of: the number of participants, \Npart (left panel), $\nu=2 \langle N_{bin} \rangle /\langle N_{part} \rangle$ (middle panel) where $N_{bin}$ is the number of binary collisions,  and the fraction of participants that undergo more than one collision, $f= \langle N_{part > 1} \rangle /\langle N_{part} \rangle$ (right panel) where $N_{part > 1}$ is the number of participants that undergo more than one collision .}
\label{fig:3}       
\end{figure*}
The left panel of figure \ref{fig:3} shows total mid-rapidity $   $ strangeness production per participant relative to p+p, where the relation in equation \ref{equ:total} is applied to both measured data and theoretical predictions, as a function of the number of participants. As expected, total strangeness follows similar trends as the singly and multi strange particles. It is worth noting that while AMPT for the current settings always under estimates strange baryon yields, it nearly always over predicts $s+\bar{s}$ (thus kaon production) yields which perhaps motivates improvements in the hadronization scheme. In particular, the introduction of baryon junctions has been shown to increase strange (anti) baryon production for other HIJING-based models \cite{BaryonJunct}.

Given the failure of  the total strangeness production to scale with the number of participants, we test to see if mid-rapidity strangeness production scales with the number binary collisions, $\langle N_{bin} \rangle$, therefore we plot $d N_{\langle s+\bar{s} \rangle}/dy$ per \Npart $ $ verses with $  $ $\nu=2 \langle N_{bin} \rangle /\langle N_{part}\rangle$ in the middle panel of figure \ref{fig:3}. Such a scaling maybe indicative of large contributions from hard processes which is shown for charm quark production at RHIC \cite{Stephen}. However, although the Cu+Cu data moved slightly closer to Au+Au data, a common trend is neither observed for the measured data nor the theoretical predictions. The middle panel of figure \ref{fig:3} also provides a test for Kharzeev-Nardi scaling of strangeness production at RHIC. This decomposes soft and hard hadron production via the following relation:
\begin{equation}
\label{equ:KNSca}
\frac{dN}{dy} = (1-x)n_{pp}\frac{N_{part}}{2} + xn_{pp}N_{bin} 
\end{equation}
where $n_{pp}$ is the p+p yield, $1-x$ is the fraction from soft processes (which scale with $N_{part}$), $x$ is the fraction from hard processes (which scale with $N_{bin}$), and $0 \leq x \leq 1$. It can be rearranged to give:
\begin{equation}
\label{equ:KNSca2}
\frac{\frac{dN}{dy}/N_{part}}{n_{pp}/2} = x(\nu-1)+1
\end{equation}
where the left term is equivalent to the value on the y-axis of figure \ref{fig:3}. In the Kharzeev-Nardi framework, $x$ should depend on the centre of mass energy ($\sqrt{s_{NN}}$) in line with QCD predictions that the relative contribution of hard processes to particle production increases with increasing center of mass energy \cite{hardprocess}. However, there is a collision system dependance ($x$ appears larger for Cu) which is not naively expected in such a framework.

Finally, in the right panel of figure \ref{fig:3} we test to see if strangeness production scales with the number of participants that undergo more than one collision, $\langle N_{part>1} \rangle$,  therefore we plot $d N_{\langle s+\bar{s} \rangle}/dy$ per \Npart $ $ verses $f=$ $ $ $\langle N_{part > 1} \rangle/\langle N_{part} \rangle$. For this scaling regime the measured data move closest together compared to the other scaling variables, while both EPOS and APMT Cu+Cu lines appear to lie on the respective Au+Au lines. The next section will discuss the potential relevance of the number of participants that undergo more than one collision for understanding strangeness production.

\section{Discussion}
\label{sec:3}

\begin{figure}
\resizebox{0.4\textwidth}{!}{%
  \includegraphics{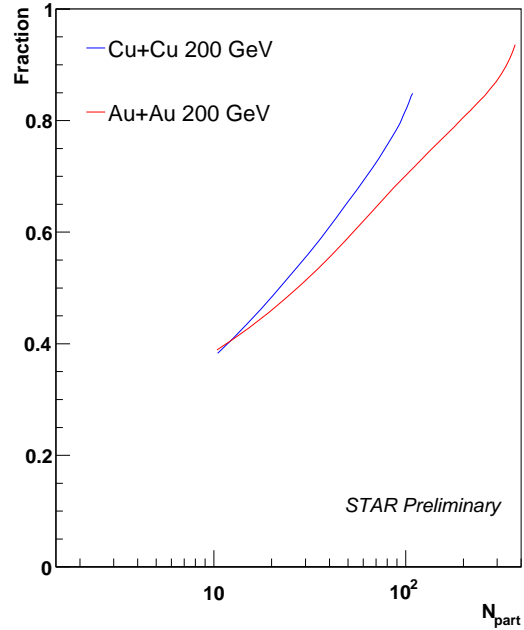}
}
\caption{Preliminary Monte Carlo Glauber calculation of the fraction of participants that undergo more than one collision as function of the number of participants for Cu+Cu and Au+Au \sNN{200} collisions.}
\label{fig:4}       
\end{figure}

Figure \ref{fig:4} shows the fraction of participants that undergo more than one collision as function of the number of participants for Cu+Cu and Au+Au \sNN{200} collisions. The distributions clearly resemble the measured data and theoretical predictions in figure \ref{fig:1} which is expected from the scaling relations in the right panel of figure \ref{fig:3}. The success of this variable in the AMPT framework can be attributed to the fact participants with more than collision have more transverse momentum transfered to the valence quarks, and when these participants decay via string breaking, they produce more strangeness compared to participants with just one collision \cite{BZhang}. As the fraction of these participants increases with centrality for a given system, strangeness production per participant also increases. When this fraction is higher in Cu+Cu compare Au+Au at a given number of participants, again strangeness production is higher for the former. 

Regarding the EPOS framework, the number of participants that undergo more than one collision is likely to have a direct relation to the size of core which, as mentioned, is the major source of strangeness production. Such a relation is expected as these participants are likely to sit in the high density region of the collision as observed in figure \ref{fig:2}. As for the fraction of participants than undergo more than one collision, it has been shown previously that core strangeness production per participant increases with the number of participants in Au+Au \cite{EPOS}. Therefore a simple explanation for higher strangeness production in Cu+Cu at a given number of participants is a larger core compared to Au+Au. This would also be inferred from the higher number of participants that undergo more than one collision in Cu+Cu.

Finally, we address how these observations possibly relate to the original hypothesis of strangeness production as a QGP signature at RHIC. In this regard, the EPOS description is closest as the core aims to represent deconfined matter with a unique hadron production scheme compared p+p. If strangeness equilibrates or indeed over-saturates in a QGP based on the original arguments, it is not inconceivable that $\gamma_{S}>=1$ upon hadronisation as mimicked by the core. On the other hand, the AMPT model offers an alternative view where soft strangeness production occurs in independently decaying nucleons rather than a coupled deconfined medium relevant to the original hypothesis. As observed in figure \ref{fig:3}, this mechanism with the final state hadron interactions appears sufficient in producing the total number of mid-rapidity strange quarks. However, as also noted, the hadronization of these strange quarks leads to insufficient hyperon yield predictions compared to the measured data where EPOS  generally does better. For further distinguishing power, it would interesting to see whether either model predicted the observed $\phi$ enhancement factor above one which is again contrary to expectations from the Canonical formalism at RHIC energies \cite{phiEnSQM}. Indeed, an alternative core-corona approach has been shown to describe the $\phi$ enhancement factors in Au+Au \sNN{200} collisions quite well \cite{phiCoreCorona}.

\section{Summary}
\label{sec:4}

In these proceedings we have shown the systematic trends for mid-rapidity strange particle yields as a function of centrality for Cu+Cu and Au+Au \sNN{200} collisions. We have found that both AMPT and EPOS models reproduce the qualitative aspects for strangeness yields per participant with EPOS doing better quantitively for the strange baryon yields. It was also shown that  despite the differing hadronization schemes, the successes of both models can be understood in the relatively simpler terms of nuclear geometry i.e. how the number of participants and the number of participants that undergo more than one collision evolve with centrality and system, which suggests this has a strong relevance to strangeness production at RHIC energies.

%
%

\end{document}